\documentclass[intlimits,twoside,a4paper]{article}

\usepackage{amsmath,amssymb}
\usepackage{graphicx,epstopdf}

\usepackage[T2A]{fontenc}
\usepackage[cp1251]{inputenc}
\DeclareMathOperator{\sech}{sech}

\usepackage[eqsecnum]{cmpj3}

\issue{2017}{20}{4}{43704}
\doinumber{10.5488/CMP.20.43704}

\title[Analysis of the effect of  polarization traps and shallow impurities]%
{Analysis of the effect of polarization traps and shallow impurities on the  interlevel light absorption of quantum dots}
\author[V.I. Boichuk, R.Ya. Leshko, D.S. Karpyn]{V.I. Boichuk, R.Ya. Leshko, D.S. Karpyn}
\address{
Department of Theoretical and Applied Physics, and Computer Simulation, 
Ivan Franko Drohobych State Pedagogical University, 
3 Stryiska St., 82100 Drohobych,  Ukraine  
}
\date{Received March 20, 2017, in final form May 27, 2017}
\begin{document}

\maketitle

\begin{abstract}
A spherical quantum dot (QD) heterosystem CdS/SiO$_2$ has been studied. Each QD has a hydrogen-like impurity in its center.
Besides that, it has been accounted that a polarization trap for electron exists at the interfaces due to the difference between the
QD and matrix dielectric permittivity. It has been defined that for  small QD radii there are surface electron states. For
different radii,  partial contributions of the surface states into the electron energy caused by the
electron-ion and electron-polarization charges interaction have been defined.
The linear light absorption coefficient of  noninteracting QDs has been calculated  taking into account the QD dispersion
by the size. It is shown that the surface states can be observed into different ranges of an electromagnetic spectrum.
\pacs 73.21.La, 78.20.Ci
\end{abstract}

\section{Introduction}

It is a long period of time that the physics of nanosystems is considered to be a priority direction in the modern development of the materials
science and  nanoelectronics. Its success is due to the use of nanoobjects such as quantum films, quantum wires,
quantum rings and especially quantum dots (QDs).
Due to their unique properties, QDs are widely used in optoelectronics. The narrow spectrum of radiation of monodisperse QDs
makes it possible to use QDs in light-emitting diodes (LEDs). They have better spectral characteristics and a higher coefficient
of efficiency than LEDs on the basis of the liquid crystals and organics materials~\cite{korbut}.

In experimental works it was defined \cite{korbut,Michalet} that the use of the nanocrystals of CdS, CdTe, CdSe makes it possible
to get an emission band in a visible spectrum. Those emission peaks are connected with exciton luminescence. Moreover, the range of size of
QDs in that work was 2--8~nm. In that and other works \cite{Kapitonov,Hoheisel,Lifei,Hasselbarth}, authors assure that except
excitonic peaks there is a wide luminescence band which is caused by the impurity and surface states.

There are many theoretical works which are devoted to the study of shallow donor and acceptor impurity states, in nanosystems of different
shapes \cite{Zhu,Polupanov,Vahdani,Boichuk1,Nasri,Xie,Boichuk2,Holovatskyi,Boichuk3}. It was proved that by decreasing  the QD radius
to the value less than the corresponding  effective Bohr radius, the probability of electron residence outside QD is larger than in the QD.
In this case, there are bound states outside the QD.

An increasing ratio of the number of atoms on the surface to the number of atoms in the nanocrystal volume (QD size becomes small)
increases the role of the surface states in the formation of absorption and luminescence bands. It relates to the heterosystem with CdS QDs.
Notwithstanding that in CdS QD it is often hard to determine the nature of the surface states,  in  most cases the red
radiation spectrum is caused by electron transition with concern to the surface traps \cite{Romcevic}. One of the reasons for the existence of a surface
trap  is the polarization charges on the interfaces. The value of polarization charges and the potential energy of the electron
interaction with polarization charges are defined by the difference of the dielectric permittivity of a heterosystem. A larger
difference of the dielectric permittivity of a heterosystem and smaller sizes of QDs  increase the role of the polarization trap.

Thus, the aim of this work is to determine the following:
\begin{itemize}
 \item the interface states of the QD with hydrogen-like impurity in the center of the QD;
 \item the effect of the surface states of the QD on the interlevel light absorption which is
 caused by the transitions between the inner and the outer (interface) states.
\end{itemize}

\section{The energy spectrum of the electron in a small QD with impurity}

Let us consider the CdS/SiO$_2$ heterosystem (table~\ref{tab1} in the appendix) consisting of a dielectric or semiconductor matrix that contains spherical QDs with a
hydrogen-like impurity in the QD center. The charged particle is characterized by its own effective mass in each medium ($m_1^*$, $m_2^*$).
The media are described by their own dielectric permittivity ($\varepsilon_1$, $\varepsilon_2$).

Modern technology is capable of obtaining a sufficient quality of  semiconductor and dielectric nanoheterostructures. In reality, it is
difficult to create a heterogeneous system with a sharp change of all physical parameters at the interface. There is always an
intermediate layer in which a particular physical parameter (particle’s effective mass,
dielectric constant) varies from its value in some crystal to the corresponding value in the other crystal.

Let us assume that at the interface there exists a transitional layer where dielectric permittivity changes from its value in the QD to the corresponding
matrix value. In this case, one may obtain the potential energy of the charge particle interaction with polarization charges as  follows
\cite{Boichuk4}:
\begin{align}
  {V_{\text p}}( r) &= \frac{\gamma }{4\varepsilon ( r)}\int\limits_0^\infty  \rd{r_0}\frac{{\tanh\left( {\frac{{{r_0} - a}}{L}} \right) + \frac{{{r_0}}}{L}\sech^2\left( {\frac{{{r_0} - a}}{L}} \right)}}{{r_0^2 - r}}
\nonumber\\
 & \quad+  \frac{\gamma^2}{8 \piup \varepsilon (r) }\int\limits_0^\infty{\rd r_0 \frac{ \frac{2r_0}{L} \sech^2\left(\frac{r_0-a}{L} \right) \tanh \left( \frac{r_0-a}{L} \right) + \tanh^2 \left( \frac{r_0-a}{L} \right)}{r_0^2-r^2}  } 
\nonumber\\
 &\quad + \frac{\gamma^2}{16 \piup \varepsilon (r) r } \int\limits_0^\infty {\rd r_0 \ln{\left|\frac{r_0+r}{r_0-r}\right|} \frac{1}{r_0} \left[ \frac{2r_0}{L} \sech^2 \left( \frac{r_0-a}{L} \right) \tanh \left( \frac{r_0-a}{L} \right) + \tanh^2 \left( \frac{r_0-a}{L} \right) \right] } 
\nonumber\\
 & \quad+ \frac{\gamma^2}{4 \piup \varepsilon ( r) r } \int \limits_0^r {{\rd r_0 \tanh\left( {\frac{{{r_0} - a}}{L}} \right) + \frac{{{r_0}}}{L}\sech^2\left( {\frac{{{r_0} - a}}{L}} \right)}}
\nonumber\\ 
 &\quad\times \int\limits_r^\infty {\rd r_1 \frac{1}{r_0^2+r_1^2} \left[ \frac{r_1}{L} \sech^2 \left( \frac{r_1-a}{L} \right) + \tanh\left( \frac{r_1-a}{L} \right) \right]},
\label{vp}
\end{align}
\begin{equation}
\label{vp_epsilon}
  \varepsilon \left( r \right) = \frac{{{\varepsilon _1} + {\varepsilon _2}}}{2}\left[ {1 - \gamma \tanh\left( {\frac{{r - a}}{L}} \right)} \right],
\end{equation}
\begin{equation}
 \label{gamma}
  \gamma  = \frac{{{\varepsilon _1} - {\varepsilon _2}}}{{{\varepsilon _1} + {\varepsilon _2}}}\,.
\end{equation}

We used Hartree units ($m_0 = 1$, $\hbar=1$, $e=1$). Based on the formula~(\ref{vp}), the analysis shows that the transitional
layer width is approximately equal to the crystal constant $a_0$,
if $L \leqslant 1/4 \cdot a_0$. $L$ is the theory parameter which determines the transition layer width. In the case of the matrix near
the QD surface, there is a potential well which is called a polarization trap.

\begin{figure}[!b]
\begin{center}
\includegraphics[width=0.6\textwidth]{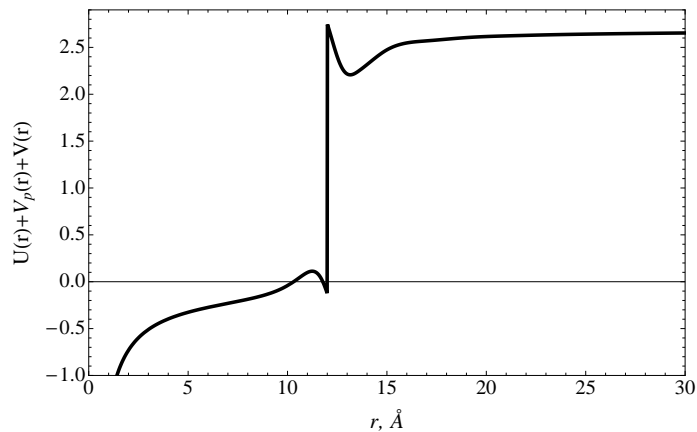}
\caption{The electron potential energy as a function of $r$ in the QD heterosystem CdS/SiO$_2$. QD radius $a=12$~\AA, $L=a_0/4$.} \label{fig1}
\end{center}
\end{figure}

We consider the electron of spherical QD CdS in the matrix SiO$_2$. The Hamiltonian of the heterosystem has the form:
\begin{equation}
 \label{hamiltonian}
  {\bf{\hat H}} =  - \frac{1}{2}\pmb\nabla \frac{1}{{m\left( r \right)}}\pmb\nabla  + U\left( r \right) + V(r) + {V_{\text p}}\left( r \right) = {{\bf{\hat H}}^0} + {V_{\text p}}\left( r \right),
\end{equation}
where the confinement potential is
\begin{equation}
 \label{confinement}
  U\left( r \right) = \left\{ \begin{array}{ll}
  0, & r \leqslant a,\\
  {U_0}\,, & r > a,
  \end{array} \right.\,\,\,\,\,{U_0} > 0,
\end{equation}
where $a$ is QD radius, and the potential energy $V_{\text p}(r)$ is expressed by~(\ref{vp}). Also, the impurity ion interacts with the electron.
The potential energy of this interaction was derived from the Poisson equation and has the form
\begin{equation}
 \label{modified_coulomb}
 V( r) = \left\{ \begin{array}{ll}
 - \displaystyle\frac{1}{{{\varepsilon _1}r}} - \frac{{{\varepsilon _1} - {\varepsilon _2}}}{{{\varepsilon _1}{\varepsilon _2}a}}\,, & r \leqslant a, \vspace{1mm}\\
 - \displaystyle\frac{1}{{{\varepsilon _2}r}}\,, & r > a.
\end{array} \right.
\end{equation}

The total electron potential energy is plotted in  figure~\ref{fig1}.

The Schr\"{o}dinger equation with Hamiltonian ${{\bf{\hat H}}^0}$ can be solved exactly \cite{Boichuk2}. Solutions
can be expressed in two different regions of $r$:
\begin{equation}
 \label{psi0}
 {\psi ^{(0)}}\left( {r,\Omega } \right) = \left\{ \begin{array}{ll}
{\psi _1}\left( {r,\Omega } \right),& r \leqslant a,\\
{\psi _2}\left( {r,\Omega } \right),& r > a,
\end{array} \right.
\end{equation}
where $\Omega$ is a solid angle.

Let $r \leqslant a$. Then,
\begin{equation}
 \label{psi0_1_e_m}
{\psi _1}\left( {{\xi _1},\Omega } \right) = {A_1}{{\mathop{M}\nolimits} _{{\lambda _1},l + 1/2}}\left( {{\xi _1}} \right)/{\xi _1} \cdot Y_l^m\left( \Omega  \right),
\end{equation}
if
\begin{displaymath}
 {\tilde E_1} = E + \frac{{\left( {{\varepsilon _1} - {\varepsilon _2}} \right)}}{{{\varepsilon _1}{\varepsilon _2}a}} < 0,
\end{displaymath}
where $\xi_1 = \alpha_1 r$, $(\alpha_1)^2 = -8 m_1^* {\tilde E_1}$, $\lambda_1=2 m_1^* / (\varepsilon_1 \alpha_1)$, $M$ is Whittaker function, or
\begin{equation}
 \label{psi0_1_e_b}
 \psi _1^{}\left( {{\xi _1},\Omega } \right) = A_1^{}{{\mathop{F}\nolimits} _l}\left( {{\delta _1},{\xi _1}} \right)/{\xi _1} \cdot Y_l^m\left( \Omega  \right),
\end{equation}
if
\begin{displaymath}
 {\tilde E_1} = E + \frac{{\left( {{\varepsilon _1} - {\varepsilon _2}} \right)}}{{{\varepsilon _1}{\varepsilon _2}a}} > 0,\,
\end{displaymath}
where ${\xi _1} = {\beta _1}r$, $\beta _1^2 = 2m_1^*{\tilde E_1}$, ${\delta _1} =  - m_1^*/\left( {{\varepsilon _1}{\beta _1}} \right)$, $F$ is a regular Coulomb function.
On condition that $r>a$
\begin{equation}
 \label{psi0_2}
 {\psi _2}\left( {{\xi _2},\Omega } \right) = {D_2}{{\mathop{W}\nolimits} _{{\lambda _2},l + 1/2}}\left( {{\xi _2}} \right)/{\xi _2} \cdot Y_l^m\left( \Omega  \right),
\end{equation}
where  ${\xi _2} = {\alpha _2}r$, ${\left( {{\alpha _2}} \right)^2} =  - 8m_2^*{\tilde E_2}$,  ${\lambda _2} = 2m_2^*/\left( {{\varepsilon _2}{\alpha _2}} \right)$,  ${\tilde E_2} = E - {U_0} < 0$, $W$ is the other
Whittaker function, which tends to zero, if $r \rightarrow \infty$.

Using the boundary conditions and a normalizing condition
\begin{eqnarray}
&{ {{R_1}\left( r \right)} \big|_{r = a}} - { {{R_2}\left( r \right)} \big|_{r = a}} = 0,&\nonumber\\
&{ \displaystyle{\frac{1}{{m_1^*}}\frac{\partial }{{\partial r}}{R_1}\left( r \right)} \Big|_{r = a}} - { {\dfrac{1}{{m_2^*}}\dfrac{\partial }{{\partial r}}{R_2}\left( r \right)} \Big|_{r = a}} = 0,&\nonumber\\
&{\int {\rd\vec r\left| {\psi \left( {r,\theta ,\varphi } \right)} \right|} ^2} = 1,&
 \label{boundary}
\end{eqnarray}
the energy spectrum of the electron is calculated and all the constants are found.

\begin{figure}[!t]
\begin{center}
\includegraphics[width=0.49\textwidth]{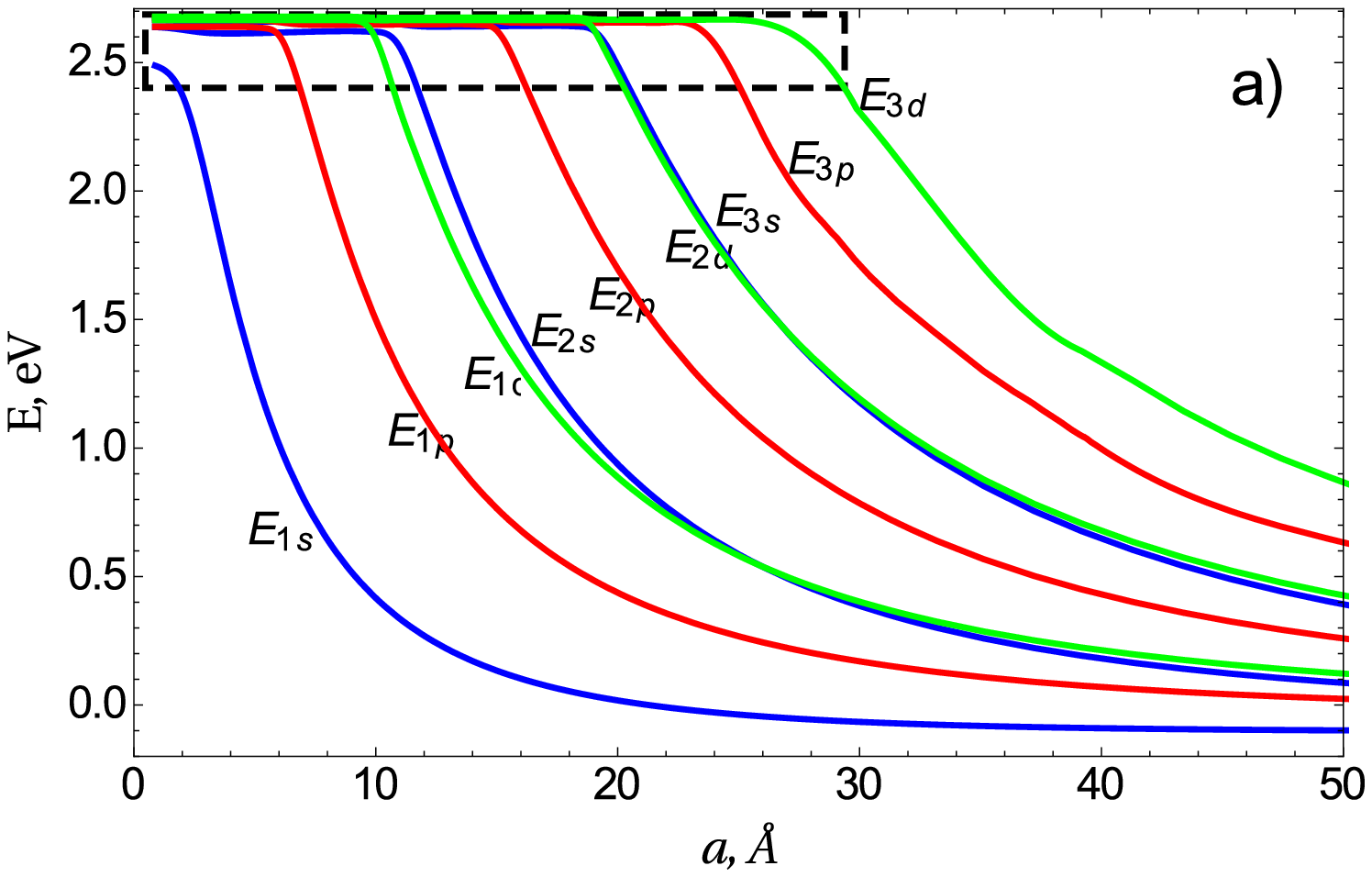}
\includegraphics[width=0.49\textwidth]{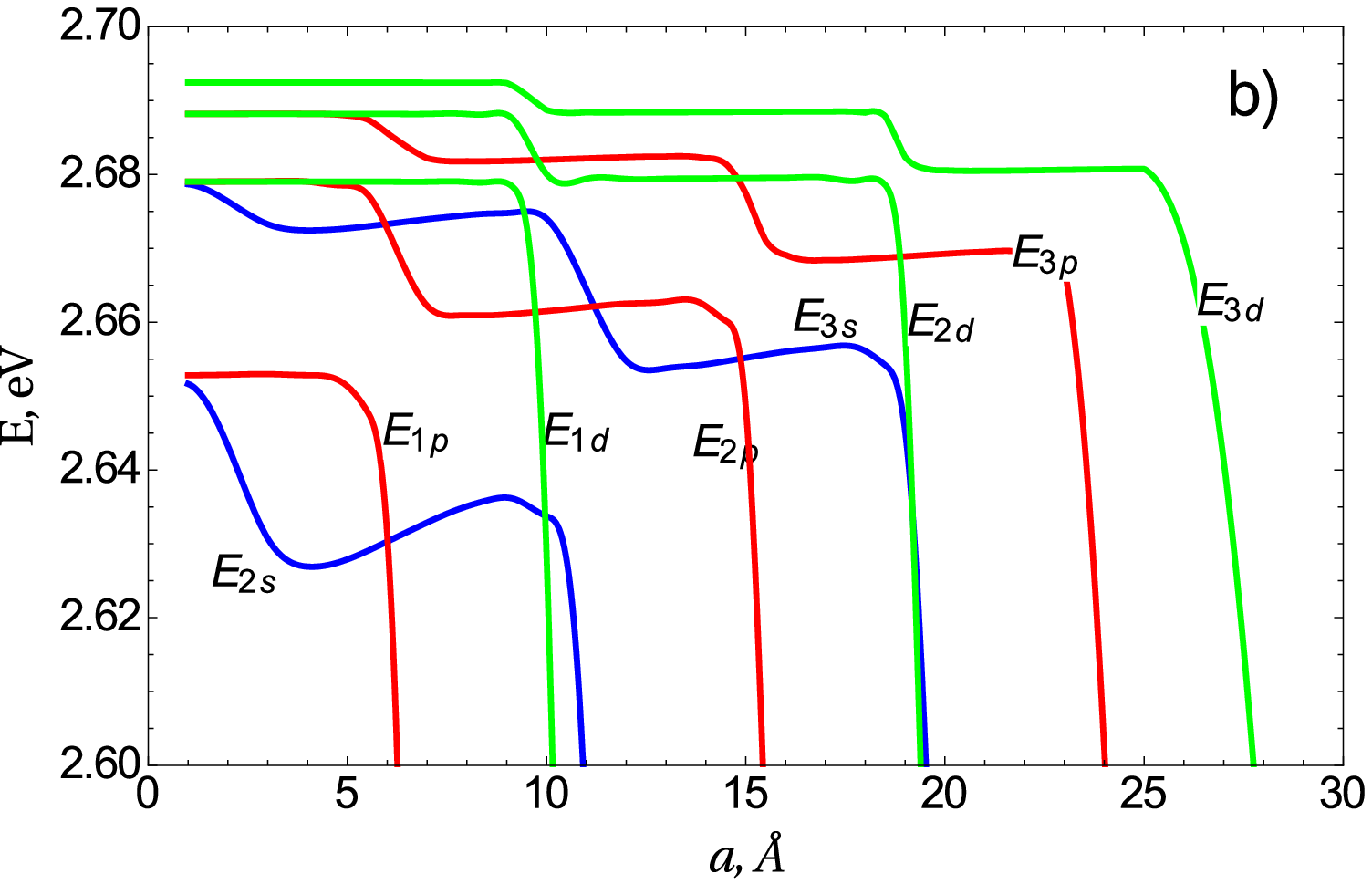}
\caption{(Color online) The electron energy in the QD. Figure (b) shows the box from figure (a). } \label{fig2}
\end{center}
\end{figure}

The calculation results of the electron for $s$-, $p$-, $d$-states were presented in figure~\ref{fig2}.
From the graphs it is seen that for  each state there is a range where a decrease of the QD radius leads
to a particle energy increase due to a space confinement.

A decrease of the QD radius shows that for some range of the radii of QD, the energy of each state reaches some values,
where it does not depend on $a$. According to the number of energy levels there are several regions of radii. These energy dependences
on the QD radius can be explained by the fact that for a specific state, the QD radius decrease is accompanied by a push of the
energy level from the quantum well. In this case, the electron energy monotonously increases. When the probability density of the particle
has its maximum in the matrix near the interface (in the surface state) its energy does not depend on the QD radius.
From figure~\ref{fig2}~(b) it is seen that each lower surface level ``pushes'' the upper level of that type. And we have got sets of levels that
correspond to hydrogen-like energy levels.

The above described results were obtained from the exact solution of the Schr\"odinger equation with~${{\bf{\hat H}}^0}$.
The surface states were caused only by the ion of impurity which keeps an electron in the matrix near the interface.
In a real heterosystem, there are polarization charges that influence  the surface states. The Schr\"{o}dinger equation with
Hamiltonian~(\ref{hamiltonian}) can be solved taking polarization charges into consideration. The wave function which is a solution
to the Schr\"{o}dinger equation with Hamiltonian~(\ref{hamiltonian}) is expressed by the expansion over the function~(\ref{psi0}):
\begin{equation}
 \label{psi}
 \Psi  = \sum\limits_i^{} {{C_i}{\psi _i}^{(0)}},
\end{equation}
$i$ is the quantum number set $n$, $l$, $m$. Our task  reduces to the linear system of equations
\begin{equation}
 \label{system_equation}
 \sum\limits_i^{} {{C_i}\left\{ {\big[ {E_i^{(0)} - E} \big]{\delta _{ji}} + \big\langle {\psi _j^{(0)}} \big|{V_{\text p}}\big| {\psi _i^{(0)}} \big\rangle } \right\}}  = 0.
\end{equation}

\begin{figure}[!t]
\begin{center}
\includegraphics[width=0.59\textwidth]{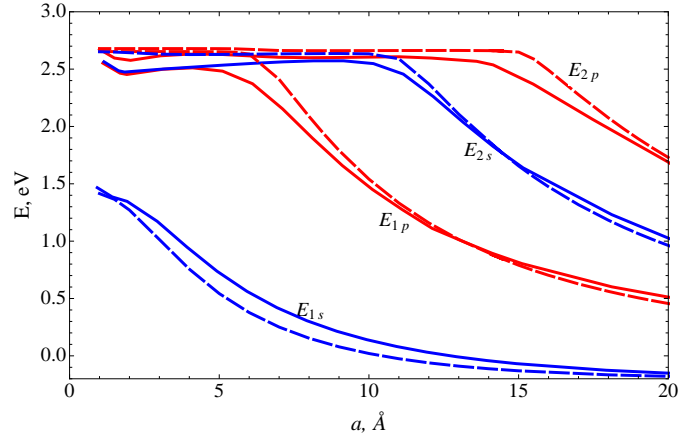}
\caption{(Color online) Impurity electron energy with regard to the polarization charges (solid curves) and with a neglect of the polarization charges (dashed curves).} \label{fig3}
\end{center}
\end{figure}

From~(\ref{system_equation}) and from the normalizing condition, all $C_i$ and the energy of the system have been found. The calculation results are
presented in figure~\ref{fig3}.

Figure~\ref{fig3} shows that  the polarization charges being taken into consideration increase the electron energy for the inner-dot states.
In this case, we receive the known result \cite{Boichuk2, Boichuk4}: if the QD dielectric permittivity is larger than the matrix one,
the QD energy  increases. However, for  small QD radii, taking the polarization charges into consideration decreases the electron energy.
This result can be understood when one sees the potential~(\ref{vp}) plotted in figure~\ref{fig1} for QD radius 12~\AA.
Figure~\ref{fig1} shows that outside  the QD, there is a potential well (polarization trap).
Obviously, at the small QD sizes, the electron gets into the polarization trap due to the confinement.
Hence, taking the polarization charges into account for  small QD radii decreases the electron energy.
Therefore, the polarization charges and the ion of impurity are two independent reasons for the existence of surface states.
As one can see from figure~\ref{fig3}, for  very small QD radii (less than 10~\AA), the results are only of theoretical interest,
but they do not contradict to the resolution  made above. The crystals exist where the above mentioned effects are observed for
 larger QD radii. This depends on the effective Bohr radius ($a_b^*$). For CdS $a_b^*=14.5$~\AA.

Let us calculate the absorption of electromagnetic waves caused by the electron interlevel transition from some $|i\rangle$ state into $|f\rangle$ state.
Let the heterosystem be irradiated by a linear polarized light. Then, in the dipole approximation, the selection rules
are $\Delta l=\pm 1$ and $\Delta m=0$. By the use of formulae from \cite{Vahdani}, we write the light absorption coefficient for the
heterosystem:
\begin{equation}
 \label{alpha_single}
 {\alpha _{i,f}}\left( \omega  \right) = \omega \sqrt {\frac{{{\mu _0}}}{{{\varepsilon _0}\varepsilon }}} \frac{{N{{| {{d_{if}}} |}^2}\hbar \Gamma }}{{{{\left( {{E_f} - {E_i} - \hbar \omega } \right)}^2} + {{\left( {\hbar \Gamma } \right)}^2}}}\,,
\end{equation}
where $\varepsilon_0$ is electric constant, $\mu_0$ is magnetic constant, $d_{if}$ is matrix element of dipole moment,
$N$ is carrier concentration. The concentration of the QDs in the matrix was chosen according to the value that provides the case
when the interaction between QDs can be neglected.
$\hbar \Gamma$ is relaxation rate. If $T \approx 4$~K, then $\hbar \Gamma \rightarrow 0$:
\begin{equation}
 \label{alpha_delta}
 {\alpha _{i,f}}\left( \omega  \right) = \mathop {\lim }\limits_{\hbar \Gamma  \to 0} \left[ {\omega \sqrt {\frac{{{\mu _0}}}{{{\varepsilon _0}\varepsilon }}} \frac{{N{{| {{d_{if}}} |}^2}\hbar \Gamma }}{{{{\left( {{E_f} - {E_i} - \hbar \omega } \right)}^2} + {{\left( {\hbar \Gamma } \right)}^2}}}} \right] = \omega \piup \sqrt {\frac{{{\mu _0}}}{{{\varepsilon _0}\varepsilon }}} N{| {{d_{if}}}|^2}\delta \left( {{E_f} - {E_i} - \hbar \omega } \right),
\end{equation}
(\ref{alpha_delta}) is valid only for a system of QDs which has an identical size.
However, in practice, the sets of QDs have a dispersion by  size. The QD size distribution can be approximated by the Lifshits-Slezov
or Gauss function. We use Gauss function:
\begin{equation}
 \label{Gauss}
 g\left( {s,\bar a,a} \right) = \frac{1}{{s\sqrt {2\piup } }}\exp \left[ { - \frac{{{{\left( {a - \bar a} \right)}^2}}}{{2{s^2}}}} \right],\
\end{equation}
where $a$ is the QD radius (variable), $s$ is half-width of the distribution~(\ref{Gauss}), which is expressed by the average
radius $\bar a$ and the value $\sigma$ which is considered as the variance in the QD sizes expressed in percentage:
$s=\bar a \sigma / 100$. By taking the QD dispersion into account (\ref{Gauss}), the absorption coefficient is obtained as follows:
\begin{equation}
 \label{alpha_system}
 {\alpha _{i,f}}\left( \omega  \right) = \omega \piup \sqrt {\frac{{{\mu _0}}}{{{\varepsilon _0}\varepsilon }}} N\sum\limits_j {\frac{{g( {s,\bar a,{a_{0j}}} )\,\,{{| {{d_{if}}( {{a_{0j}}} )} |}^2}}}{{{{\left| {\frac{\rd}{{\rd a}}\left[ {{E_f}\left( a \right) - {E_i}\left( a \right) - \hbar \omega } \right]} \right|}_{a = {a_{0j}}}}}}}\,,
\end{equation}
where $a_{0j}$ are simple zeros of the function $Q(a)=E_f(a)-E_i(a)-\hbar \omega$.

\begin{figure}[!t]
\begin{center}
\includegraphics[width=0.6\textwidth]{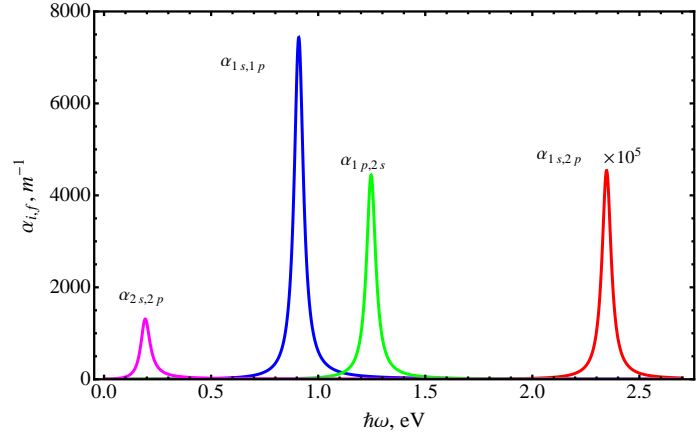}
\caption{(Color online) Light absorption coefficient caused by the electron interlevel transition.} \label{fig4}
\end{center}
\end{figure}

The dependence of the absorption coefficient on the quant energy of light for an average radius 11.64~{\AA}
(two crystals parameters $a_0$, $N \approx 3 \cdot 10^{16}$~cm$^{-3}$)
and dispersion $\sigma=5$\% is shown in  figure~\ref{fig4}. Figure~\ref{fig4} shows that the light absorption coefficient is caused
by the transition from the ground state into the surface state (1$s$-2$p$) and is 5--6 orders less than the transition from the other inner-dot states into exited states.
The reason for this is the distance between the energy levels, which is larger than in
the case of transition (1$s$-1$p$) while the dipole momentum of interlevel transition is smaller.

Furthermore, the transition from the ground state into the first exited state (1$s$-1$p$) for this value is larger and is located by the energy scale far
from the next possible transition  from the ground state to the exited 2$p$-state (1$s$-2$p$).

\section{Conclusion}
The calculation of the electron energy in the QD heterosystem with an impurity shows that the reduction of the QD size causes a
transformation of the lowest exited states from the inner-dot into the outer-dot states. The outer-dot states are characterised by a weak
energy dependence on the QD radius. The probability density of the electron in the space for those states has a maximum value near the QD
surface in the matrix. Those states are called ``surface electron states''.

Taking into account the polarization charges and the intermediate layer with $\varepsilon = \varepsilon(r)$ on the surface, causes the
existence of the electron polarization trap in the matrix near the QD surface. It has been defined that the electron interaction
with polarization
charges increases the binding energy of the surface states (decreases the electron energy). For example, for $a=15$~\AA, the
electron binding energy
for the surface 2$p$-state equals 94~meV if the potential~(\ref{vp}) is neglected, and 267~meV if the potential~(\ref{vp}) is taken into account.

Based on the found wave functions of the inner-dot and outer-dot states, we calculate the light absorption coefficient for QD
heterosystem CdS/SiO$_2$. With an average QD radius $\bar a = 2 a_0$ ($a_0$ is the crystal parameter) there are absorption bands which have
different energies ($E_{2s{\text -}2p} \approx 0.29$~eV, $E_{1s{\text -}1p} \approx 0.92$~eV, $E_{1p{\text -}2s} \approx 1.28$~eV, $E_{1s{\text -}2p} \approx 1.37$~eV)
and heights.

The obtained results should be taken into consideration while analysing  the experimental absorption and luminescence bands for a heterosystem with
CdS QD.

\appendix
\section*{Appendix}
\phantom{.}
\renewcommand{\arraystretch}{1.2} 
\begin{table}[!h]
\vspace{-9mm}
\begin{center}
\caption{\label{tab1} Parameters of the heterosystem \cite{Holovatskyi2}.}
\vspace{2ex}
 \begin{tabular}{|c | c | c | c | c|}
 \hline\hline
 crystal & $m$ ($m_e$) & $\varepsilon$ & $a_0$, {\AA} & $U_0$, eV\\
 \hline\hline
 CdS & 0.2 & 5.5 & 5.818 & 0 \\
 \hline
 SiO$_2$ & 0.42 & 3.9 & $-$ & 2.7 \\
 \hline\hline
\end{tabular}
\end{center}
\end{table}

\newpage

\ukrainianpart

\title{Аналіз впливу поляризаційних пасток і мілких домішок на міжрівневе поглинання світла квантовими точками}
\author{В.І. Бойчук, Р.Я. Лешко, Д.С. Карпин}
\address{
Кафедра теоретичної і прикладної фізики та комп'ютерного моделювання, 
Дрогобицький державний педагогічний університет імені Івана Франка, 
вул. Стрийська, 3, 82100 Дрогобич,  Україна 
}

\makeukrtitle

\begin{abstract}
\tolerance=3000%
У роботі досліджується наногетеросистема сферичних квантових точок (КТ) CdS у матриці SiO$_2$. Кожна КТ у центрі містить
водневоподібну домішку. Крім того, враховано, що біля меж поділу через різницю діелектричних проникностей виникає поляризаційна
пастка для електрона. Встановлено, що для малих радіусів КТ існують поверхневі електронні стани. Для різних радіусів КТ обчислено
парціальні внески в енергію зв'язку поверхневих станів від взаємодії електрона з іоном домішки та поляризаційними зарядами.
Проведено обчислення коефіцієнта міжрівневого поглинання гетеросистеми з невзаємодіючими КТ, враховуючи їх дисперсію за
розмірами. Показано, що поверхневі стани проявляються в різних областях спектру.
\keywords  донорна домішка, коефіцієнт поглинання, поляризаційна пастка

\end{abstract}


\begin{thebibliography}{99}
\bibitem{korbut} Korbutyak~D.V., Kovalenko~O.V., Budzulyak~S.I., Kalytchuk S.M., Kupchak I.M., Ukr. J. Phys. Reviews, 2012, \textbf{7}, No.~1, 48--95 (in Ukrainian).
\bibitem{Michalet} Michalet~X., Pinaud~F.F., Bentolila~L.A., Tsay~J.M., Doose~S., Li~J.J., Sundaresan~G., Wu~A.M., Gambhir~S.S.,  Weiss~S., Science, 2005, \textbf{307}, 538, 
    \bibdoi{10.1126/science.1104274}.
\bibitem{Kapitonov} Kapitonov~A.M., Stupak~A.P., Gaponenko~S.V., Petrov~E.P., Rogach~A.L., Eychm\"uller~A., J. Phys. Chem. B, 1999, \textbf{103}, 10109,
    \bibdoi{10.1021/jp9921809}.
\bibitem{Hoheisel} Hoheisel~W., Colvin~V.L., Johnson~C.S., Alivisatos~A.P., J. Chem. Phys., 1994, \textbf{101}, 8455,
    \bibdoi{10.1063/1.468107}.
\bibitem{Lifei} Xi L., Lek J.Y., Liang Y.N., Boothroyd C., Zhou W., Yan Q., Hu X., Chiang F.B.Y., Lam Y.M., Nanotechnology, 2011, \textbf{22}, 275706,
    \bibdoi{10.1088/0957-4484/22/27/275706}. 
\bibitem{Hasselbarth} H\"asselbarth~A., Eychm\"uller~A.,  Well~H., Chem. Phys. Lett., 1993, \textbf{203}, 271, 
    \bibdoi{10.1016/0009-2614(93)85400-I}.
\bibitem{Zhu} Zhu~J.-L., Chen~X., Phys. Rev. B, 1994, \textbf{50}, 4497,
    \bibdoi{10.1103/PhysRevB.50.4497}.
\bibitem{Polupanov} Polupanov~A.F., Galiev~V.I., Novak~M.G., Fiz. Tekh. Poluprovodn., 1997, \textbf{31}, 1375 (in Russian).
\bibitem{Vahdani} Vahdani~M.R.K., Rezaei~G., Phys. Lett. A, 2009, \textbf{373}, No.~34, 3079,
    \bibdoi{10.1016/j.physleta.2009.06.042}.
\bibitem{Boichuk1} Boichuk~V.I., Bilynskyi~I.V., Leshko R.Ya., Condens. Matter Phys., 2010, \textbf{13}, 13702,
    \bibdoi{10.5488/CMP.13.13702}.
\bibitem{Nasri} Nasri~D., Sekkal~N., Physica E, 2010, \textbf{42}, 2257,
    \bibdoi{10.1016/j.physe.2010.04.028}.
\bibitem{Xie} Xie~W., Superlattices Microstruct., 2010, \textbf{48}, 239,
    \bibdoi{10.1016/j.spmi.2010.04.015}.
\bibitem{Boichuk2} Boichuk~V.I., Bilynskyi~I.V., Leshko~R.Ya., Turyanska~L.M., Physica E, 2011, \textbf{44}, 476, \\
    \bibdoi{10.1016/j.physe.2011.09.025}.
\bibitem{Holovatskyi} Holovatsky~V.A., Frankiv~I.B., J. Phys. Stud., 2012, \textbf{16}, 1706 (in Ukrainian).
\bibitem{Boichuk3} Boichuk~V.I., Bilynskyi~I.V., Leshko~R.Ya., Turyanska~L.M., Physica E, 2013, \textbf{54}, 281, \\
    \bibdoi{10.1016/j.physe.2013.07.003}.
\bibitem{Romcevic} Romcevic~M., Romcevic~N., Kostic~R., Klopotowski~L., Dobrowolski~W.D., Kossut~J., \v Comor~M.I., J. Alloys Compd., 2010, \textbf{497}, 46,
    \bibdoi{10.1016/j.jallcom.2010.03.072}.
\bibitem{Boichuk4} Boichuk~V.I., Kubai~R.Yu., Fiz. Tverd. Tela, 2001, \textbf{43}, 226 (in Russian).
\bibitem{Holovatskyi2} Holovatsky~V.A., Makhanets~O.M., Voitsekhivska~O.M., Physica E, 2009, \textbf{41}, 1522, \\$  $
    \bibdoi{10.1016/j.physe.2009.04.027}.  

\end{thebibliography}
\end{document}